\documentstyle[prb,aps,amsfonts,preprint,graphicx]{revtex}

\tightenlines
\begin{document}
\draft
\author{Zuowei Wang, Christian Holm}
\address{Max-Planck-Institut f\"ur Polymerforschung,
  Ackermannweg 10, D-55128 Mainz, Germany }
\title{Estimate of the Cutoff Errors in the \\ Ewald Summation for
  Dipolar Systems}
\date{\today}
\maketitle

\newcommand{\erfc}{{\mbox{erfc}}}
\newcommand{\infd}{{\mbox{d}}}
\newcommand{\mod}{{\mbox{{~}mod{~}}}}

\newcommand{\MM}{{\Bbb{M}}}
\newcommand{\NN}{{\Bbb{N}}}
\newcommand{\RR}{{\Bbb{R}}}
\newcommand{\ZZ}{{\Bbb{Z}}}

\newcommand{\VECk}{{\mathbf{k}}}
\newcommand{\VECl}{{\mathbf{l}}}
\newcommand{\VECm}{{\mathbf{m}}}
\newcommand{\VECn}{{\mathbf{n}}}
\newcommand{\VECr}{{\mathbf{r}}}
\newcommand{\VECz}{{\mathbf{z}}}
\newcommand{\VECE}{{\mathbf{E}}}
\newcommand{\VECF}{{\mathbf{F}}}
\newcommand{\VECR}{{\mathbf{R}}}
\newcommand{\VECmu}{{\mbox{\boldmath$\mu$}}}
\newcommand{\VECtau}{{\mbox{\boldmath$\tau$}}}
\newcommand{\VECchi}{{\mbox{\boldmath$\chi$}}}

\newcommand{\CALP}{{\mathcal P}}
\newcommand{\CALI}{{\mathcal I}}
\newcommand{\CALL}{{\mathcal L}}
\newcommand{\CALM}{{\mathcal M}}
\newcommand{\CALQ}{{\mathcal Q}}
\newcommand{\CALR}{{\mathcal R}}
\newcommand{\CALT}{{\mathcal T}}

\newcommand{\exa}{{^{\mbox{\footnotesize exa}}}}
\newcommand{\maxi}{{_{\mbox{\scriptsize max}}}}
\newcommand{\Mesh}{{_{\mbox{\scriptsize M}}}}
\newcommand{\nd}{{^{\mbox{\footnotesize nd}}}}
\newcommand{\opt}{{_{\mbox{\footnotesize opt}}}}
\newcommand{\self}{{^{\mbox{\scriptsize self}}}}
\newcommand{\th}{{^{\mbox{\footnotesize th}}}}

\newcommand{\D}{\displaystyle}


\begin{abstract}
 
 Theoretical estimates for the cutoff errors in the Ewald summation
method for dipolar systems are derived. Absolute errors in the total energy, 
forces and torques, both for the real and reciprocal space
parts, are considered. The applicability of the estimates is tested
and confirmed in several numerical examples. We demonstrate that these 
estimates can be used easily in determining the optimal parameters of 
the dipolar Ewald summation in the sense that they
minimize the computation time for a predefined, user set, accuracy.

\end{abstract}
\vspace{2ex}

\narrowtext

%
%

\section{Introduction}
%
Computer simulations are becoming more and more important in the
study of complex systems. Among them many of the most interesting systems are
charged systems, and as such naturally dominated by long range interactions.
This is certainly the case for almost all biological systems where the
electrostatic interactions play a dominant role\cite{holm01a}, but there is
also a wealth of technological important substances such as polyelectrolytes, where the
charged nature is one of the key ingredient for their functionality
\cite{hara93a}. The electrostatic interactions can be of monopolar origin, like
interactions between proteins, DNA, or charged membranes, but also of dipolar
nature, because all biological tissues contain water, which is a dipolar
substance.  The simultaneous appearance of dipoles and monopoles is of great
importance, for example, for protein folding simulations\cite{gunsteren99a}.
Dipolar interactions can also be of technological importance, for example in
the application of so called ferrofluids, which are basically dispersed
magnetic particles\cite{Rosensweig85a}. However, computer simulations
of long range interactions in periodic boundary conditions are
notoriously difficult to handle, and possess an unfavorable scaling
with the amount of particles involved. The most often used method to
compute the interactions relies on the famous Ewald
sum\cite{ewald21a}.  In the simplest implementations the involved
computation time grows like $N^2$, or at best like $N^{3/2}$, if the
cutoff is optimally varied with the splitting parameter\cite{perram88a}. The
use of fast Fourier transformations (FFT) can further reduce
the scaling to basically $N\cdot \log N$. There have been quite a
number of advances in the application of these so-called
particle-mesh-Ewald techniques for Coulomb systems over the last few
years\cite{hockney88a,darden93a,essmann95a,deserno98a,huenenberger00a}. An
important aspect of all algorithms is the tuning in the sense of speed at well
controlled errors. All algorithms have parameters such as the real space
cutoff $r_c$, the reciprocal space cutoff $k_c$, the splitting parameter $\alpha$,
and for the mesh based methods there are even more. To find the optimal
combination and at the same time control the systematic error is a formidable
task, and cannot efficiently be achieved by trial and error. For the
simulations of point charges, there have been reliable error estimates
published for the standard Ewald summation \cite{kolafa92a}, and for the mesh Ewald
methods PME \cite{petersen95a} and P3M \cite{deserno98b}. For the dipolar
Ewald  summation, there has been just an estimate given for the energy
in real-space in Ref.\onlinecite{kolafa92a}. For molecular dynamics (MD) simulations of dipolar
systems, however, we need to know the errors for the forces and the torques. In
this article we give a reliable error estimate for
the energy, for the forces, and for the torques, when computed via the standard
Ewald sum. We will show the applicability of these estimates by comparing them
to well specified systems.  Moreover we will give a detailed discussion on the
optimization of the parameters, which will lead to the most efficient
parameters for a predefined error in each observable quantity. This can all be done
prior to the actual simulation, ensuring thus optimal performance at optimally
controlled errors.

The article is organized as follows: In Sec. II we briefly review the
important formulas of the Ewald summation for dipolar systems. The
theoretical estimates of the cut-off errors in the Ewald sum are derived
in Sec. III. In Sec. IV the estimates are compared with numerical
accuracy measurements for specified model systems, and they are found
to be very precise. The use of the estimates in determining 
the optimal parameters and their application
to an inhomogeneous system is discussed in Sec. V and VI, respectively.
Finally we end with some conclusions in Sec. VII.
%

\section{The Ewald Summation} 
%
%
Consider a system of $N$ particles with a point-dipole $\VECmu_{i}$ 
at their center position ${\VECr}_i$ in a cubic simulation box of 
length $L$. If periodic boundary conditions are applied, 
the total electrostatic energy of the box is given by
\begin{equation} \label{Unormal}
U =\frac{1}{2}\sum_{i=1}^{N}\sum_{j=1}^{N} \sum_{\VECn \in \ZZ^3}^{~~~~\prime}
\bigg{\{} \frac{\VECmu_i \cdot \VECmu_j}{\mid \VECr_{ij}+\VECn \mid^3} -
\frac{3[\VECmu_i \cdot (\VECr_{ij}+\VECn)][\VECmu_j\cdot
(\VECr_{ij}+\VECn)]}{\mid \VECr_{ij}+\VECn \mid ^{5}} \bigg{\}}. 
\end{equation}    
where ${\VECr}_{ij}={\VECr}_i-{\VECr}_j$. The sum over $\VECn$ is 
over all simple cubic lattice points, $\VECn=(n_xL, n_yL, n_zL)$ with
$n_x, n_y, n_z$ integers. The prime indicates that the $i=j$ term must
be omitted for $\VECn=0$. The slowly decaying long range part of the 
dipolar potential renders the straightforward summation of Eqn.(\ref{Unormal})
impracticable. The Ewald method provides an
efficient way of calculating $U$, which splits the problem into two
absolutely and rapidly convergent parts, one in real space and one 
in reciprocal space. The details of the method are discussed in Refs.
\onlinecite{ewald21a,perram88a,allen87a,deleeuw80a}, here we only give the final expressions 
\begin{equation}
 U=U^{(r)}+U^{(k)}+U^{(self)}+U^{(surf)},
\end{equation}
where the real-space $U^{(r)}$, the k-space (reciprocal-space) $U^{(k)}$,
the self $U^{(self)}$ and the surface $U^{(surf)}$ contributions are
respectively given by:  
\begin{eqnarray}
U^{(r)}&=&\frac{1}{2}\sum_{i=1}^{N}\sum_{j=1}^{N}\sum_{{\VECn} \in
  \ZZ^3}^{~~~~\prime} \bigg\{({\VECmu}_i \cdot
{\VECmu}_j) B(\mid {\VECr}_{ij}+{\VECn}\mid)- [{\VECmu}_i \cdot
({\VECr}_{ij}+\VECn)][{\VECmu}_j \cdot ({\VECr}_{ij}+\VECn)] C(\mid {\VECr}_{ij}
+{\VECn}\mid)\bigg\},\label{energyrr}\\
U^{(k)}&=&\frac{1}{2 L^3}\sum_{\VECk \in \ZZ^3,{\VECk} \ne
  0}\frac{4\pi}{k^2}\exp[-(\pi k/\alpha L)^2] \sum_{i=1}^{N}\sum_{j=1}^{N}
 ({\VECmu}_i \cdot {\VECk})({\VECmu}_j \cdot {\VECk})
\exp(2 \pi i {\VECk} \cdot {\VECr}_{ij}/L),\label{energykk}\\
U^{(self)}&=& -\frac{2\alpha^3}{3\sqrt{\pi}}\sum_{i=1}^{N}\mu_i^2,\label{energyself}\\
U^{(surf)}&=&\frac{2\pi}{(2\epsilon^{\prime}+1)L^3}\sum_{i=1}^{N}\sum_{j=1}^{N}{\VECmu}_i
\cdot {\VECmu}_j.\label{energysurf}
\end{eqnarray}
Here the sums over $i$ and $j$ are for the particles in the central
box and
\begin{eqnarray}
 B(r)&=&[\erfc(\alpha r)+(2\alpha r /\sqrt{\pi})\exp(-\alpha^2
 r^2)]/r^3,\label{functionB}\\
 C(r)&=&[3\erfc(\alpha r)+(2\alpha r/\sqrt{\pi})(3+2\alpha^2 r^2)
        \exp(-\alpha^2 r^2)]/r^5.\label{functionC}
\end{eqnarray}
The prime on the third sum in Eqn.(\ref{energyrr}) also denotes that the
divergent terms $i=j$ for ${\VECn}=0$ have to be omitted,
 $\erfc(x):=2\pi^{-1/2}\int_{x}^{\infty}
 \exp(-t^2) dt$ is the complementary error function.
The inverse length $\alpha$ is the splitting parameter of the Ewald
summation which should be chosen so as to optimize the performance.
The form Eqn.(\ref{energysurf}) given for the surface correction 
assumes that the set of the periodic replications of the simulation
box tends in a spherical way towards an infinite cluster and
that the medium outside this sphere is an uniform dielectric with
dielectric constant $\epsilon^{\prime}$ \cite{allen87a,deleeuw80a}. The case of a
surrounding vacuum corresponds to $\epsilon^{\prime}=1$ and
the surface term vanishes for the metallic boundary conditions
($\epsilon^{\prime}=\infty$). 

 In practical calculations, the infinite sums in
Eqns.(\ref{energyrr}, \ref{energykk}) are truncated
by only taking into account distances which are smaller than some real
space cutoff $r_c$ and wave vectors with a modulus smaller than some
reciprocal space cutoff $k_c$. If $r_c \le L/2$, the sum in real space
(Eqn.(\ref{energyrr})) reduces to the normal minimum image convention. The double
sum over particles in $U^{(k)}$ can be replaced by a product of two
single sums which is more suitable for numerical calculations.

The force ${\VECF}_i$ acting on particle $i$ is obtained by
differentiating the potential energy $U$ with respect to ${\VECr}_i$,
i.e.,
\begin{equation}
{\VECF}_i=-\frac{\partial U }{\partial {\VECr}_i}={\VECF}_i^{(r)}+{\VECF}_i^{(k)},
\end{equation}
with the real-space and k-space contributions given by:
\begin{eqnarray}
{\VECF}^{(r)}_i & = & \sum_{j=1}^{N} \sum_{\VECn\in\ZZ^{3}}^{~~~~\prime}\Bigg{\{}\{
({\VECmu}_i \cdot {\VECmu}_j)(\VECr_{ij}+\VECn)+{\VECmu}_i [{\VECmu}_j 
\cdot ({\VECr}_{ij}+\VECn)]+[{\VECmu}_i \cdot
({\VECr}_{ij}+\VECn)]{\VECmu}_j \} C(\mid {\VECr}_{ij}+{\VECn}\mid)
\nonumber \\ ~ &-&[{\VECmu}_i \cdot ({\VECr}_{ij}+\VECn)][{\VECmu}_j \cdot
({\VECr}_{ij}+\VECn)] D(\mid {\VECr}_{ij}+{\VECn}\mid)({\VECr}_{ij}+{\VECn})
\Bigg{\}},\label{forceir}\\
{\VECF}^{(k)}_i&=&\frac{2\pi}{L^4}\sum_{j=1}^{N} \sum_{\VECk \in \ZZ^3,{\VECk} \ne
  0}\frac{4\pi{\VECk}}{k^2} ({\VECmu}_i \cdot {\VECk}) ({\VECmu}_j \cdot {\VECk})
\exp[-(\pi k/ \alpha L)^2] \sin(2 \pi {\VECk} \cdot {\VECr}_{ij}/L),\label{forceik}
\end{eqnarray}
where
\begin{equation}
D(r)=[15\erfc(\alpha r)+(2\alpha r/\sqrt{\pi})(15+10 \alpha^2 r^2+4
\alpha^4 r^4)\exp(-\alpha^2 r^2)]/r^7.
\end{equation}
Since the self and surface energy terms [Eqns.(\ref{energyself},
\ref{energysurf})] are independent of the particle
positions, they have no contributions to the force, unlike the
Ewald summation for the Coulomb systems where the surface term contributes.

The torque $\VECtau_i$ acting on particle $i$ is related to the
electrostatic field $\VECE_i$ at the location of this particle, 
\begin{equation}
\VECtau_i=\VECmu_i \times \VECE_i=\VECtau_i^{(r)}+\VECtau_i^{(k)}+
\VECtau_i^{(surf)},
\end{equation} 
with 
\begin{equation}
{\VECE}_i=-\frac{\partial U }{\partial {\VECmu}_i},
\end{equation}
and thus
\begin{eqnarray}
\VECtau_i^{(r)}&=&-\sum_{j=1}^{N} \sum_{{\VECn} \in
  \ZZ^3}^{~~~~\prime} \bigg\{({\VECmu}_i \times
{\VECmu}_j) B(\mid {\VECr}_{ij}+{\VECn}\mid)- [{\VECmu}_i \times
({\VECr}_{ij}+\VECn)][{\VECmu}_j \cdot ({\VECr}_{ij}+\VECn)] C(\mid {\VECr}_{ij}
+{\VECn}\mid)\bigg\}, \label{torqueir}\\
\VECtau_i^{(k)}&=&-\frac{1}{L^3}\sum_{j=1}^{N} \sum_{\VECk \in \ZZ^3,{\VECk} \ne
  0}\frac{4\pi}{k^2}({\VECmu}_i \times {\VECk})({\VECmu}_j \cdot
{\VECk})  \exp[-(\pi k/ \alpha L)^2] \exp(2 \pi i {\VECk}
 \cdot {\VECr}_{ij}/L), \label{torqueik}\\
\VECtau_i^{(surf)}&=&-\frac{4\pi}{(2\epsilon^{\prime}+1)L^3}\sum_{j=1}^{N}{\VECmu}_i
\times {\VECmu}_j.\label{torqueisurf}
\end{eqnarray}

When performing computational simulations, it is important to
control the accuracy properly. In molecular dynamics methods
the accuracy is generally estimated from the root mean square (rms) error
in the forces:\cite{deserno98a,petersen95a,deserno98b}
\begin{equation} \label{rmsforce}
\Delta F 
:= \sqrt{\frac{1}{N}\sum_{i=1}^{N}\left(\Delta\VECF_{i}\right)^{2}}
:= \sqrt{\frac{1}{N}\sum_{i=1}^{N}
\left(\VECF_{i}-\VECF^{\mbox{\footnotesize exa}}_{i}\right)^{2}}
\end{equation}
where $\VECF_{i}$ is the force on particle $i$ calculated by the 
algorithm under investigation (here the Ewald sum) and 
$\VECF^{\mbox{\footnotesize exa}}_{i}$ 
is the {\em exact} force on that particle. In next section, we will
derive estimates for the rms error $\Delta F$ caused by cutting off the
Ewald summation in real-space and k-space. The similar estimates
are also derived for the cutoff errors in the total energy and
torques. There are no errors involved in the self and surface contributions
(Eqns(\ref{energyself}, \ref{energysurf}, \ref{torqueisurf})),
because no cutoff operations are applied to them.
%
\section{Explicit derivation of Error formulas}
%
\subsection{Real-Space Errors}
  
 The real-space cutoff error in the force on particle $i$ can be
written as\cite{deserno98b} 
\begin{equation}\label{simpdf}
\Delta {\VECF}^{(r)}_i  = \mid\VECmu_i\mid \sum_{j\not=i}
\mid\VECmu_j\mid \VECchi_{ij}^{(r)}.
\end{equation} 
The idea behind this is that the error on $\VECF_i^{(r)}$
originates from the $N-1$ interactions of particle $i$
with the other dipolar particles, and each contribution should be
proportional to the product of the two dipole moments involved.
The vector $\VECchi_{ij}^{(r)}$ gives the direction and magnitude 
of the error contribution from two unit dipoles in the simulation box,
depending on their separation and orientations. It follows
from Eqn.(\ref{forceir}) that $\VECchi_{ij}^{(r)}$ is given by
\begin{eqnarray}\label{eqchiij}
\VECchi_{ij}^{(r)} & = & \sum_{\VECr,r>r_c} 
\Bigg\{[\hat{\VECr}\cos\vartheta(\hat{\VECmu}_i,\hat{\VECmu}_j)+
\hat{\VECmu}_i\cos \vartheta(\hat{\VECmu}_j,\hat{\VECr})+
\hat{\VECmu}_j\cos\vartheta(\hat{\VECmu}_i,\hat{\VECr})]rC(r)
\nonumber \\ &-& \hat{\VECr} \cos \vartheta(\hat{\VECmu}_i,
\hat{\VECr})\cos \vartheta(\hat{\VECmu}_j, \hat{\VECr})r^3 D(r) \Bigg\}
\end{eqnarray}
where $\hat{\VECmu}_i$ and $\hat{\VECmu}_j$ are unit vectors along the
dipole orientations on particles $i$ and $j$, $\hat{\VECr}$ is the unit
vector along $\VECr$, and
the sum in (\ref{eqchiij}) runs over all the periodic images of
particle $j$ for $\mid \VECr:=\VECr_{ij}+\VECn\mid >
r_c$. 
$\vartheta(\hat{\VECmu}_i,
\hat{\VECmu}_j)$ denotes the angle between the vectors
$\hat{\VECmu}_i$ and $\hat{\VECmu}_j$, $\vartheta(\hat{\VECmu}_i,
\hat{\VECr})$ the angle between $\hat{\VECmu}_i$ and
$\hat{\VECr}$, and so forth. 

 To derive the estimate of $\Delta {\VECF}^{(r)}_i$, we
assume that the positions and dipole moment 
orientations of the particles separated by distances larger than $r_c$
are distributed randomly. This assumption is certainly not
valid for all dipolar systems, but it is reasonable as a starting point. 
For random systems, the error contributions from different
particles can be assumed to be uncorrelated \cite{deserno98b}
\begin{equation} \label{uncorre}
\left\langle\VECchi_{ij}^{(r)}\cdot\VECchi_{im}^{(r)}\right\rangle = 
\delta_{jm}\left\langle \VECchi_{ij}^{(r)2} \right\rangle =:
\delta_{jm} \chi^{(r)2},
\end{equation}    
where the angular brackets denote the average over all particle
configurations. Obviously, the term $\left\langle \VECchi_{ij}^{(r)2}
\right\rangle$ - the mean square force error for two unit dipoles -
can no longer depend on $i$ and $j$ and is thus written as
$\chi^{(r)2}$. Using Eqns.(\ref{simpdf}, \ref{uncorre}),
it follows that
\begin{equation} \label{deltaF1}
\left\langle (\Delta\VECF_{i}^{(r)})^{2} \right\rangle = 
\mu_{i}^{2} \sum_{j\not=i}\sum_{m\not=i} \mid \VECmu_{j}\mid\mid\VECmu_{m}\mid
\langle \VECchi_{ij}^{(r)}\cdot\VECchi_{im}^{(r)}\rangle \approx
\mu_{i}^{2} \CALM^2 \chi^{(r)2},
\end{equation}
where the quantity $\CALM$ is defined as
\begin{equation}
\CALM^2 := \sum_{j=1}^{N} \mu_{j}^{2}.
\end{equation}
To obtain the configurational average value $\left\langle \Delta F \right
\rangle$, we further assume that
\begin{equation}
  \label{eq:random}
  \left\langle\sqrt{\frac{1}{N}\sum_{i=1}^{N}\left(\Delta\VECF_{i}\right)^{2}}\right\rangle
  \approx
\sqrt{\frac{1}{N}\sum_{i=1}^{N}\langle\left(\Delta\VECF_{i}\right)^{2}}\rangle,
\end{equation}
which can be shown true for reasonably large systems by the law of large
numbers along the line of reasoning of Ref.\onlinecite{deserno98b}. Inserting
Eq.(\ref{deltaF1}) into Eqs.(\ref{rmsforce}, \ref{eq:random}) we end up with the relation  
\begin{equation}
\Delta F^{(r)} \approx \chi^{(r)}\frac{\CALM^{2}}{\sqrt{N}}.
\end{equation}

Using Eqn.(\ref{eqchiij}), it follows
\begin{eqnarray} \label{eqchi2}
\chi^{(r)2} &\approx& \frac{1}{L^3}\int_{r_c}^{\infty}r^2dr\int_{0}^{\pi}
\sin\theta d\theta \int_{0}^{2\pi}d\phi \Bigg{\{}
[\hat{\VECr}\cos\vartheta(\hat{\VECmu},\hat{\VECmu}^{\prime})+
\hat{\VECmu}\cos \vartheta(\hat{\VECmu}^{\prime},\hat{\VECr})+
\hat{\VECmu}^{\prime}\cos\theta]rC(r) \nonumber \\ &-& \hat{\VECr}
\cos \theta\cos \vartheta(\hat{\VECmu}^{\prime}, \hat{\VECr})r^3 D(r)\Bigg{\}}^2,
\end{eqnarray}
where $\hat{\VECmu}$ and $\hat{\VECmu}^{\prime}$ denote the unit
orientation vectors of two arbitrary dipoles. The z-axis of the
spherical coordinates ($r$, $\theta$, $\phi$) is chosen to be along the
orientation $\hat{\VECmu}$. That gives $\vartheta(\hat{{\VECmu}}, \hat{\VECr}) 
=\theta$ and 
\begin{equation}\label{triangle}
\cos \vartheta(\hat{\VECmu}^{\prime},\hat{\VECr})= \cos\theta \cos\vartheta
(\hat{\VECmu}, \hat{\VECmu}^{\prime})+\sin\theta \sin\vartheta(\hat{\VECmu},\hat
{\VECmu}^{\prime}) \cos\phi.
\end{equation} 
In the integrand of Eqn.(\ref{eqchi2}), all the triangular functions
about $\vartheta(\hat{{\VECmu}}, \hat{{\VECmu}}^{\prime})$ should be
replaced by their configuration average values. This means that
the cross terms which contain odd powers of
$\cos/\sin\vartheta(\hat{{\VECmu}},\hat{{\VECmu}}^{\prime})$
will vanish due to their zero mean value. The remaining
terms $\cos^2/\sin^2 \vartheta(\hat{{\VECmu}},\hat{{\VECmu}}^{\prime})$ 
are replaced by their mean values of $1/2$. Following the asymptotic
expansion formula \cite{kolafa92a}
\begin{equation} \label{asym}
\int^{\infty}_{A} \exp(-B x^2) f(x) dx \approx \exp(-BA^2)f(A)/2BA
\end{equation}
which is valid when $B>0$ and $d[f(x)/2Bx]/dx \ll f(x)$ for $x \geq A$,
the complementary error function is approximated by
\begin{equation} \label{ercfapp}
\erfc(\alpha r) \approx \exp(-\alpha^2 r^2)/\sqrt{\pi} \alpha r. 
\end{equation}
Eqn.(\ref{ercfapp}) is introduced into (\ref{eqchi2}) through 
(\ref{functionB}, \ref{functionC}). Estimating the integral
again with Eqn.(\ref{asym}) gives the approximate expression
of $\chi^{(r)2}$. We obtain
\begin{equation}
\chi^{(r)2} \approx L^{-3} r_c^{-9}\alpha^{-4} (\frac{13}{6}C_c^{2}+\frac{2}{15}
D_c^2-\frac{13}{15} C_c D_c) \exp(-2 \alpha^2 r_c^2),
\end{equation}
where the terms $C_c$ and $D_c$ are given by
\begin{eqnarray}
C_c &=& 4 \alpha^4 r_c^4+6\alpha^2 r_c^2+3,\\
D_c &=& 8\alpha^{6} r_c^6+20 \alpha^4 r_c^4+ 30 \alpha^2 r_c^2+15.
\end{eqnarray}
The resulting rms expectation of the real-space cutoff error
in the forces is thus 
\begin{equation} \label{forcer}
\Delta F^{(r)} \approx \CALM^2  (L^{3}\alpha^4 r_c^{9} N)^{-1/2}
(\frac{13}{6}C_c^{2}+\frac{2}{15}
D_c^2-\frac{13}{15} C_c D_c)^{1/2} \exp(-\alpha^2 r_c^2).
\end{equation}
The derivation of the expected real-space cutoff errors in the total
potential energy and torques follows the same way. 
For calculating the fluctuation of the error in total energy, it is 
noted that the interaction energy between two dipoles is
evenly shared between them. That means the sum of
$\langle (\Delta U^{(r)})^2 \rangle$  over all particles contains
each pair contribution twice and thus the fluctuation of the
real-space cutoff error is one half of the sum \cite{kolafa92a}.
Then the rms value of the real-space cutoff error
of the total potential energy is estimated as
\begin{equation} \label{energyr}
\Delta U^{(r)} \approx \CALM^2 (L^{3} \alpha^{4} r_c^{7} N)^{-1/2} 
[\frac{1}{4}B_c^2 + \frac{1}{15}C_c^2-\frac{1}{6}B_c C_c]^{1/2}
\exp(-\alpha^2 r_c^2)
\end{equation}
with
\begin{equation}
B_c=2\alpha^2 r_c^2+1.
\end{equation}
The rms error on the torques is estimated similarly to the force as
\begin{equation} \label{torquer}
\Delta \tau^{(r)} \approx \CALM^2 (L^{3} \alpha^{4} r_c^{7} N)^{-1/2} 
[\frac{1}{2}B_c^2 + \frac{1}{5}
C_c^2 ]^{1/2} \exp(-\alpha^2 r_c^2).
\end{equation}
Eqn.(\ref{forcer}, \ref{energyr}, \ref{torquer}) all contain the
exponential $\exp(-\alpha^2 r_c^2)$. For sufficiently low errors,
$\alpha r_c$ has to be larger than one, for example $\alpha r_c \approx
\pi$ for an error of $\exp(-\pi^2) \approx 5 \times 10^{-5}$.
If only the highest powers of $\alpha r_c$ are retained,
the estimates of the real-space cutoff errors in the
total energy, forces and torques can be reduced to
\begin{eqnarray} 
\Delta U^{(r)} &\approx& 4 \CALM^2 \alpha^2  (r_c/15 NL^3)^{1/2} 
\exp(-\alpha^2 r_c^2),\label{energyrs}\\
\Delta F^{(r)} &\approx& 8 \CALM^2 \alpha^4 (2r_c^3/ 15NL^3)^{1/2} 
\exp(-\alpha^2 r_c^2),\label{forcers}\\
\Delta \tau^{(r)} &\approx& 4 \CALM^2 \alpha^2(r_c/5NL^3)^{1/2} 
\exp(-\alpha^2 r_c^2), \label{torquers}
\end{eqnarray}
where Eqn.(\ref{energyrs}) is a factor of $\sqrt{6/5}$ slightly larger than
that given in Eqn.(35) of Ref.\onlinecite{kolafa92a}. The advantage of these
simplified formulas is that they reflect the dependence of the rms
errors on $\alpha$ and $r_c$ more directly and thus could be used 
more easily in determining the optimal values of these parameters. 

\subsection{Reciprocal-Space Errors}
 
In deriving the estimates of the reciprocal-space (k-space)
cutoff errors, we further assume that the radial distribution function 
of the particles is approximately unity at all distances. Following
Eqn.(\ref{forceik}), the k-space cutoff error in the force acting on
particle $i$ is given by
\begin{equation}\label{sumfk}
\Delta {\VECF}_{i}^{(k)}=\sum_{j=1}^{N}\sum_{\VECk,k>k_c}\frac{8\pi^2{\VECk}}
{L^4 k^2} ({\VECmu}_i \cdot {\VECk})({\VECmu}_j \cdot {\VECk})
\exp[-(\pi k/ \alpha L)^2] \sin (2 \pi {\VECk} \cdot
{\VECr}_{ij}/L).
\end{equation}
Note that the diagonal term ($j=i$) in the sum does not depend on the
positions of the particles. It will provide a systematic contribution to
the cutoff error in k-space \cite{kolafa92a}. In Eqn.(\ref{sumfk}) this contribution 
equals to zero, thus there is no systematic part of the error in the
forces. The same thing happens to the error in the torques. But for
the total energy the diagonal terms are positive and the systematic
contribution plays a dominant role in the cutoff error.

  The off-diagonal terms in Eqn.(\ref{sumfk}) do depend on the
positions of the particles and have alternating signs. The statistical
approach in Sec. III.A can also be used. Similar to Eqn.(\ref{simpdf}),
the off-diagonal contribution to the cutoff error in
$\Delta {\VECF}_{i}^{(k)}$ is given by
\begin{equation}
\Delta {\VECF}^{(k)}_{i,off}  = \mid\VECmu_i\mid \sum_{j\not=i}
\mid\VECmu_j\mid \VECchi_{ij}^{(k)}
\end{equation} 
with
\begin{equation} \label{chik1}
\VECchi_{ij}^{(k)} = \sum_{\VECk,k>k_c}\frac{8\pi^2\VECk}{L^4}
\cos\vartheta (\hat{\VECmu}_i,\hat{\VECk}) \cos\vartheta
(\hat{\VECmu}_j,\hat{\VECk}) \exp[-(\pi k/ \alpha L)^2] 
i \exp(2 \pi i {\VECk} \cdot {\VECr} / L),
\end{equation}
where $\VECr$ stands for $\VECr_{ij}$ and $\sin(2 \pi {\VECk} 
\cdot {\VECr}/L)$ is re-written as $i \exp(2 \pi i {\VECk} \cdot
{\VECr}/L)$ according to the symmetrical character of the summation
over $\VECk$. $\hat{\VECk}$ is the unit vector along $\VECk$.
Since the particles are 
assumed to be randomly distributed over the simulation box, the
fluctuation of $\Delta\VECF_{i,off}^{(k)}$ can also be written as
\begin{equation} \label{deltafk}
\left\langle (\Delta\VECF_{i,off}^{(k)})^{2} \right\rangle = 
\mu_{i}^{2} \sum_{j \neq i}\sum_{m \neq i} \mid \VECmu_{j}\mid
\mid\VECmu_{m}\mid \langle \VECchi_{ij}^{(k)}\cdot\VECchi_{im}^{(k)\ast}
\rangle \approx \mu_{i}^{2} \CALM^2 \chi^{(k)2}.
\end{equation}
Again  $\chi^{(k)2}$ is independent of $i$ and $j$.
Using Eqn.(\ref{chik1}), we have
\begin{equation} \label{chik2}
\chi^{(k)2} \approx {\bigg{(}\frac{8\pi^2}{L^4}\bigg{)}}^2   
\int_{k_c}^{\infty} \exp[-2(\pi k/ \alpha L)^2] k^4 dk \int_{0}^{\pi}
\sin\theta d\theta \int_{0}^{2\pi} \cos^{2}\vartheta (\hat{\VECmu},
\hat{\VECk}) \cos^{2}\vartheta(\hat{\VECmu}^{\prime},\hat{\VECk}) d\phi.
\end{equation} 
Choosing the z-axis of the spherical coordinates ($k, \theta, \phi$)
along the $\hat{\VECmu}$ orientation, the same discussion as in
Eqn.(\ref{eqchi2}) gives 
\begin{equation}
\chi^{(k)2} \approx 128 \pi^3 L^{-6} \alpha^2 k_c^3  \exp[-2(\pi k_c/ 
\alpha L)^2]/15. 
\end{equation}  
The rms expectation of the k-space cutoff error in the forces 
is thus  
\begin{equation} \label{forcek}
\Delta F^{(k)} \approx 8 \pi \CALM^2 L^{-3} \alpha (2\pi k_c^3/15N)^{1/2}
\exp[-(\pi k_c/ \alpha L)^2].
\end{equation}
Here the notation $\Delta F_{off}^{(k)}$ is replaced directly with
$\Delta F^{(k)}$ due to the fact of no diagonal contribution.

The derivation of the off-diagonal parts of the cutoff errors in the
total energy and torques proceeds in the same way. That the sum over
$\langle (\Delta U_{i,off}^{(k)})^2 \rangle$ contains each pair
contribution twice has also been considered in the error estimate
of the total energy. The results are given by
\begin{eqnarray}
\Delta U^{(k)}_{off} &\approx& 4 \CALM^2 L^{-2} \alpha (\pi k_c/15N)^{1/2}
\exp[-(\pi k_c/ \alpha L)^2], \label{offuk}\\
\Delta \tau^{(k)} &\approx&  4 \CALM^2 L^{-2} \alpha (\pi k_c/5N)^{1/2}
\exp[-(\pi k_c/ \alpha L)^2]. \label{torquek}
\end{eqnarray} 
$\Delta \tau^{(k)}$ is also used directly instead of $\Delta 
\tau^{(k)}_{off}$.

 The diagonal (systematic) part of the cutoff error in the total energy can be
written as
\begin{eqnarray}\label{ukdiag}
\Delta U^{(k)}_{diag} &=& \frac{1}{2\sqrt{N}} \sum_{i=1}^{N}\sum_{\VECk,k>k_c}
\frac{4 \pi}{L^{3}} \mu_i^{2} \cos^{2} \vartheta(\hat{\VECmu}_i,\hat{\VECk})
\exp[-(\pi k/ \alpha L)^2] \label{ukoff} \nonumber \\
~&\approx& \frac{2 \pi}{L^{3} \sqrt{N}} \CALM^2 \int_{k_c}^{\infty}
\exp[-(\pi k/ \alpha L)^2] k^2 dk \int_{0}^{\pi} \sin \theta d\theta
\int_{0}^{2\pi} \cos^2\vartheta(\hat{\VECmu},\hat{\VECk}) d\phi \nonumber \\
~&\approx& \frac{4}{3} \CALM^2 L^{-1} \alpha^2 k_c N^{-1/2}
 \exp[-(\pi k_c/ \alpha L)^2],
\end{eqnarray}
where the sum is again approximated by an integral and 
the asymptotic expansion formula of Eqn.(\ref{asym}) is used to
get the final estimate. The total k-space cutoff error in the total
energy is thus
\begin{equation} \label{energyk}
\Delta U^{(k)}=\Delta U^{(k)}_{diag}+\Delta U^{(k)}_{off}.
\end{equation}
Comparing Eqn.(\ref{ukdiag}) with (\ref{offuk}), it can be seen that
the systematic part of the error is a factor of $\sim L \alpha
k_c^{1/2} (\gg 1)$ larger than the statistical part. Hence the systematic
contribution is dominant in the k-space cutoff error of the total energy. 

Assuming that the real-space and reciprocal-space contributions to the 
error are independent, the total cutoff error in Ewald summation can be written as
\begin{equation}
\Delta \Theta =\sqrt{\Delta \Theta^{(r)^2}+\Delta \Theta^{(k)2}},
\end{equation}
where $\Theta$ stands for $U, F$ and $\tau$. 
%
\section{Comparison of Formula with numerical calculation}
%

 In this section, we carry out numerical experiments to check 
the validity of the error estimates derived in the previous section.

  In order to make our measurements fully reproducible, we choose the test
system to be consistent with the one described
in Appendix D of a previous publication:\cite{deserno98a} 100 particles randomly
distributed within a cubic box of length $L=10$, each of them has a 
unit point-dipole at the center. The coordinates and dipole
orientations of the 100 particles are constructed by generating
500 random numbers $\CALR_n$ ($n=1,\cdots 500$) between $0$ and $1$.
The first 300 random numbers give the Cartesian coordinates of the 
particles as ($x_i=L\CALR_{3i-2},y_i=L \CALR_{3i-1},z_i=L\CALR_{3i}$).
The remaining 200 numbers distribute the dipole orientations uniformly over a
unit sphere's surface by the following way:
\begin{eqnarray}
\cos\theta_i &=& 2\CALR_{300+2i-1}-1.0,\\
\phi_i &=& 2\pi \CALR_{300+2i},
\end{eqnarray}
yielding ($\hat{\VECmu}_{ix} =\sin\theta_i\cos\phi_i,\hat{\VECmu}_{iy}
 = \sin\theta_i\sin\phi_i, \hat{\VECmu}_{iz} = \cos\theta_i$). 
The function $\tt rand$ which can be found in many $\tt C$ libraries
is used as the random number generator. Thus the positions of the
particles are exactly the same as that discussed in
 Refs.\onlinecite{deserno98a,deserno98b}. 
Our unit conventions are as follows: lengths are measured in $\CALL$
and dipoles in $\CALP$. Hence the unit of the energy, force and torque
are $\CALP^2/\CALL^3,\CALP^2/\CALL^4$ and $\CALP^2/\CALL^3$, respectively.

 The {\em exact} energy, forces and torques are obtained by performing
a direct summation in real space with Eqn.(\ref{Unormal}) and the
related derivatives. The sum over $\VECn$ is built up in a spherical way 
up to $\mid \VECn_{cut} \mid = 40 L$. The Ewald summation calculations
are carried out under vacuum condition
($\epsilon^{\prime}=1$) in order to compare the results with
the direct summation\cite{deleeuw80a,nymand2000a}. The rms errors
in the forces and torques are evaluated directly with the definition 
of Eqn.(\ref{rmsforce}). However, it is inconvenient to do this for the
energy due to the existence of the constant contributions ($U^{(self)}$ and
$U^{(surf)}$), so the rms error in the total energy is simply taken as 
\begin{equation} \label{rmsenergy}
\Delta U = \left \langle (U - U^{\mbox{\footnotesize
      exa}})^2 \right \rangle^{1/2},
\end{equation}
where the angular brackets denote the configuration average.
If only one specific random system is considered, the
use of Eqn.(\ref{rmsenergy}) may make the result on $\Delta U$ somewhat
sensitive to the details of the generated configuration. But it
is easy to see that this sensitivity can be diminished by taking
an average over several configurations.

  In the first step, we fix the real-space cutoff to $r_c = 5.0$ $(L/2)$.
The k-space cutoff $k_c$ changes from $2$ to $12$. The resulting
curves for the rms errors in the forces and torques are plotted in
Fig.1(a) and (b) together with the analytical estimates derived in Sec. III. 
To avoid the unfavorable sensitivity to configuration details, the
result on the rms error of the total energy [Fig.1(c)] is taken from an average
over $10$ random configurations which are generated one after the other
in the same way as the model system. Fig.1(a-c) show that for each $k_c$ 
there exists an optimal value of $\alpha$ which gives the
minimum rms errors. For smaller values of $\alpha$, the error
contribution from the real-space is dominant, while for larger values
the k-space contribution dominates. It can be clearly seen that
the analytical estimates accurately predict both the real-space
and k-space contributions to $\Delta F$, $\Delta \tau$ and  $\Delta U$
for all values of $\alpha$. Only for very small values of $k_c$
deviations are observed for large $\alpha$. As will be shown in the next section,
this permits an easy way to determine the optimal parameters for a
predefined accuracy. The discrepancy in the k-space part at very low
$k_c$ is to be expected as the replacement of the summation over
$\VECk$ with an integral in Eqn.(\ref{chik2}) turns to be a rather
crude approximation at this time. Since the optimal values of $k_c$
for the Ewald method are ranging between $7$ and $25$ \cite{petersen95a}, this
discrepancy does not affect the validity of the analytical estimates.
As accuracy on the energy may be obscured by unimportant constant
contributions and is sensitive to fluctuations, the rms errors on the
forces and torques are more suitable for the accuracy measurements in 
MD simulations. Furthermore, since Fig.1(a, b) show the similar behavior
of $\Delta F$ and $\Delta \tau$ as function of $\alpha$ and $k_c$ (for
example, they give the same optimal value of $\alpha$ for each $k_c$),
we will concentrate on discussions of $\Delta F$ in the remaining
part of this paper.  The same kind of results, however, have been
achieved for $\Delta U$ and $\Delta \tau$ in all the following cases.
 
  As most discussions are focused on the k-space error contributions
in Fig.1 by changing $k_c$, we fix $k_c$ in the next step and change
different values for $r_c$ in order to further investigate
the accuracy of the analytical estimates in predicting
the real-space error contributions. The same model system as
in Fig.1(a, b) is studied. $k_c$ is fixed to 8 and $r_c$ is taken
to be $3.0, 3.6, 4.3$ and $5.0$, respectively. For comparison, both 
Eqn.(\ref{forcer}) and
(\ref{forcers}) are used to estimate $\Delta F^{(r)}$. The results
in Fig.2 show that the former formula precisely gives the real-space
contribution to $\Delta F$ for all values of $r_c$.  As
expected, Eqn.(\ref{forcers}) underestimates $\Delta F^{(r)}$ at small
values of $\alpha$. Since Eqns.(\ref{energyrs}-\ref{torquers}) are
obtained under the assumption of $\alpha r_c \gg 1$, the valid
range of these simplified formulas shifts to larger $\alpha$ with 
decrease of $r_c$. However, if an accuracy as $\delta \le 5\times
10^{-5}$ is required in a MD simulation so that $\alpha$ could not
be very small considering the practical limitation of $r_c \leq L/2$,
these estimates are almost as accurate as the full formulas
[Eqns.(\ref{forcer}), (\ref{energyr}) and (\ref{torquer})].
 
  The last step is to demonstrate that the scaling of the rms errors
with the particle number and dipole moment distribution is correctly
given by $\Delta F \propto \CALM^2 N^{-1/2}$. Three
random systems which differ only in the values of $\CALM^2$ and $N$ are
investigated. The first system is the same as that studied in Fig.1-2.
The second system contains 200 particles among which 100 have a dipole
strength of $\CALP$ and the other 100 have $3 \CALP$.  The third
one contains 400 particles: 100 with $\CALP$, 200 with 
$5 \CALP$ and 100 with $7 \CALP$. Hence their values of $(\CALM^2,N)$
are respectively given by (100, 100), (1000, 200) and (10000, 400), and
the ratio of their prefactors is thus $1 : \sqrt{50} : 50$. The
results of $\Delta F$ in Fig.3 clearly reflect this scaling behavior
by the constant shift of the three curves with respect to each other
(note the logarithmic scale in the vertical axis). The analytical
estimates predict the rms errors very precisely in all the three
cases.
%
\section{Optimization of parameters}
%
 In this section, we discuss the use of the analytical 
formulas derived in Sec. III to determine the optimal values 
of $\alpha$, $r_c$ and $k_c$ by which the 
required accuracy could be satisfied and the computation time is 
minimized. The detailed discussions on this subject can also be found in
Refs.\onlinecite{perram88a,petersen95a}. 
  
  The overall computation time for computing the forces with
the Ewald method is approximately given by \cite{petersen95a}
\begin{equation} \label{time}
\CALT = {\it a_{r}} N^2 (r_c/L)^3 + {\it a_{k}}N k_c^3,
\end{equation}
where the primitive overheads ${\it a_{r}}$ and ${\it a_{k}}$
highly depend on the implementation of the code and need to be found 
by numerical experiments. As an example, we have 
carried out the time experiments on a DEC personal workstation (CPU 433MHz) 
using a standard Fortran 77 compiler. In the implementation the
complementary error function and its derivative are
calculated with table lookup and the reciprocal-space
summation is optimized as in Refs.\onlinecite{perram88a,allen87a}. The linked-cell
method is used to deal with the short-range forces (when doing simulations).
The primitive overheads are then found roughly to be $a_{r} = 2.5 \mu s$ and
$a_{k} = 0.7 \mu s$.

 For a required accuracy $\delta$, the parameters $\alpha$, $r_c$ and $k_c$
should be chosen to minimize $\CALT$ with respect to
the two constraints of the error bounds [Eqns.(\ref{forcer})
and (\ref{forcek})], which are restated as
\begin{eqnarray}
\frac{\delta}{\sqrt{2}} &=& \CALM^2  (L^{3}\alpha^4 r_c^{9} N)^{-1/2}
(\frac{13}{6}C_c^{2}+\frac{2}{15} D_c^2-\frac{13}{15} C_c D_c)^{1/2} 
\exp(-\alpha^2 r_c^2), \label{restater} \\ 
\frac{\delta}{\sqrt{2}} &=& 8 \pi \CALM^2 L^{-3} \alpha (2\pi
k_c^3/15N)^{1/2} \exp[-(\pi k_c/ \alpha L)^2]. \label{restatek}
\end{eqnarray} 
In case of $\delta \le 5\times 10^{-5}$, Eqn.(\ref{forcers}) could be
used instead of (\ref{forcer}) so as to show the dependence of the
accuracy on the parameters more clearly. Eqn.(\ref{restater}) and 
(\ref{restatek}) provide the qualitative
function relations of $r_c$ and $k_c$ with $\alpha$ as:
$r_c(\alpha) \approx - A \sqrt{\ln \delta}/\alpha$ and 
$k_c(\alpha) \approx - B \sqrt{\ln \delta} \alpha $.
Inserting them into Eqn.(\ref{time}) and differentiating it with respect
to $\alpha$ yields $\alpha \propto N^{1/6}$ and thus $r_c \propto
N^{-1/6}$ and $k_c \propto N^{1/6}$. The minimized computation time 
is then proportional to $N^{3/2}$ with the proportionality constant
depending on the accuracy. The same results can be found for
the Coulomb Ewald method in Refs.\onlinecite{perram88a,kolafa92a,petersen95a}. This can be easily 
understood by comparing Eqns.(\ref{forcer}) and
(\ref{forcek})in Sec. III of this paper with Eqns.(18) and (32) in
Ref.\onlinecite{kolafa92a} and finding the same exponential dependences of the
cutoff errors on  $\alpha$, $r_c$ and $k_c$
for the dipolar and Coulomb Ewald summations.

 The numerical investigation of the functional dependence of the optimal
parameters on $N$ and $\delta$ are performed by using the primitive
overheads obtained above. For each given $N$ and $\delta$, we at first
choose different values for $r_c$ within the inequality $r_c \le
L/2$. For each $r_c$ the parameters $\alpha$ and $k_c$ are
calculated by solving Eqns.(\ref{restater}) and (\ref{restatek}).
These values are then introduced into Eqn.(\ref{time}) to
figure out the optimal value of $r_c$ which gives the minimum
computation time. In calculations the size of the simulation cell
is fixed to a dimensionless length of $L = 10$. The range of accuracy requirement
and number of particles are chosen to be $\delta=10^{-2}$ to $10^{-5}$
measuring in $\CALP^2/\CALL^4$  and $N = 10^3$ to $N = 10^6$ which
should cover most of the applications. The particles are supposed 
to have an uniform dipole moment of $\CALP$. The results for the optimal values
of the parameters and the corresponding computation time per particle 
are shown in Fig. 4(a-d), respectively. It can be clearly seen that 
the functional dependence of the parameters and the overall
computation time on $N$ are just as discussed above. Fig.4(c) shows
that when a high accuracy is required for a system with a small number 
of particles, the predicted real-space cutoff is larger than half of
the box length and $r_c = L/2$ must be used. The optimal $\alpha$ 
values hardly depend on the accuracies. These results
are very similar to that obtained for the Coulomb Ewald summation
in Ref.\onlinecite{petersen95a}, except for $r_c \propto
N^{-1/6}$ here and $r_c \propto N^{1/6}$ there. This is
because they considered a system of constant density, while we choose
the volume of the simulation cell to be constant.
%
\vspace{-1cm}
\section{Application To An Inhomogeneous System}
%
So far we have only used the homogeneous random systems to test the
error estimates. This is of course not always the case in  
computer experiments. Many simulations will encounter
the problem of highly nonuniform distributions of particles.    
In this section we use an anisotropic dipolar system 
to investigate the influence of the inhomogeneity on the 
rms errors.

  The dipolar system we studied is an electrorheological (ER) fluid
consisting of spherical dielectric particles with uniform diameter $\sigma$
dispersed in a solvent. In the initial configuration $200$ particles
are randomly distributed in a cubic simulation box with side length of 
$10\sigma$, which gives a typical volume fraction of particles as
$\rho\approx0.1$. Upon application of an external electric field, the
dielectric particles are polarized. Each particle has an uniform
dipole moment along the field direction (z-axis) at its center under the
'point-dipole' approximation. The electrostatic interaction between 
the particles will draw them to form chain or cluster structures along
the field direction. This structuring process of the ER system 
in the quiescent state is studied by the Brownian dynamic simulation
method described in our previous paper \cite{wangzuo2000}.
The final configuration, which is used for the rms error calculation,
is obtained after a long enough simulation time when there is no
obvious structure evolution in the system. The formation of several
thick chain-like structures has been observed in the snapshot of the
system. It is also reflected clearly by the appearance of the sharp
peaks at the positions of $r=\sigma, \sqrt{3}\sigma, 2\sigma, \sqrt{7}\sigma,
3\sigma \cdots$ in the radial distribution function (RDF) $g_{0}(r)$
of the final configuration shown in Fig.5, where the RDF of the
initial configuration is also plotted for comparison.
The rms force error $\Delta F$ and the 
corresponding estimates [Eqns.(\ref{forcer}) and (\ref{forcek})] for 
the final configuration are given in Fig.6. A deviation of the
estimates from the numerical results could be observed. At smaller
values of $\alpha$ the prediction of the estimates is smaller than
that given by the algorithm, while at large values of $\alpha$ it is larger than
the actual one. By studying the derivation
of the error estimates, this could be understood qualitatively by the
reason that the formation of the chain structures leads to higher
local density of the particles and thus results in higher rms errors in the
real-space part \cite{petersen95a}. Though this effect also exists in the k-space
part, the well ordering, or in other word well spacing, of particles
along the z-direction leads to an enhanced k-space accuracy. This could
be partially seen from Eqn.(\ref{forceik}) where only the z-component $k_z$ 
involves in the calculation due to the z-direction orientation of the dipole
moments in this case. 

From the numerical curve in Fig.6, the optimal splitting parameter 
is given to be $\alpha \approx 0.72\sigma^{-1}$ with a corresponding
$\Delta F \approx 1.74 \times 10^{-5}$, while the intersection point
of the real and reciprocal space estimates occurs at $\alpha \approx 
0.71\sigma^{-1}$, which predicts an error of $\Delta F\approx
1.90\times 10^{-5}$ by Eqn.(\ref{restater}) and (\ref{restatek}).
If the estimated value of $\alpha$ is used, this would result in an
error of $\Delta F\approx 1.88\times 10^{-5}$, which is about $8\%$
larger than that at the calculated optimal value of $\alpha$. If such
a safety margin has already been considered at the beginning of the 
simulation, the determination of the optimal parameters discussed 
in the previous section is still a good approach. In addition,
it should be noted that the structure studied in Fig.6 is an extreme 
case in the simulations of ER fluids or similar systems such as 
magnetorheological (MR) fluids and ferrofluids. When the thermal
agitation is comparable with the dipolar interactions, or even more
a shear flow is introduced, the percolating chains or columns will be
broken into clusters. The distribution and orientation of the smaller clusters
would have more random characters compared with that studied in this
section, and thus a smaller discrepancy between the estimates and
actual numerical results could be expected. 
In any case, if it is suspected that the development of strongly 
inhomogeneities in the systems may lead to potential failure 
of the presented error formulas, some simple numerical tests as
performed above could provide valuable informations.
%
\section{Conclusion}
%
 The analytical formulas for the cutoff errors in the
Ewald summation for dipolar systems have been derived
in closed form. Errors in the total energy, forces and 
torques are all considered. The high quality of these
estimates has been proven in different random
systems, and thus provides an easy way to determine the optimal
tuning parameters which can give the expected accuracy, but 
minimize the computation time. Based on this,
the functional dependence of the optimal splitting parameter
$\alpha$, real-space cutoff $r_c$ and reciprocal-space 
cutoff $k_c$ on the number of particles $N$ are discussed
qualitatively and confirmed numerically by the timing experiments.
Although the validity of the error estimates is subject to
some additional requirements, such as the homogeneity of the system,
a consultation of these formulas and a priori estimate of the optimal
parameters should always be a good starting point in using the Ewald
summation for simulations of dipolar systems.

\section{Acknowledgments}

We thank H. W. M\"{u}ller and B. D\"{u}nweg for helpful discussions.
Financial support from the DFG is gratefully acknowledged.


%
%
%

\clearpage


\begin{figure}[htb]
 \begin{center}
  \includegraphics*[height=15.0 cm, angle=270]{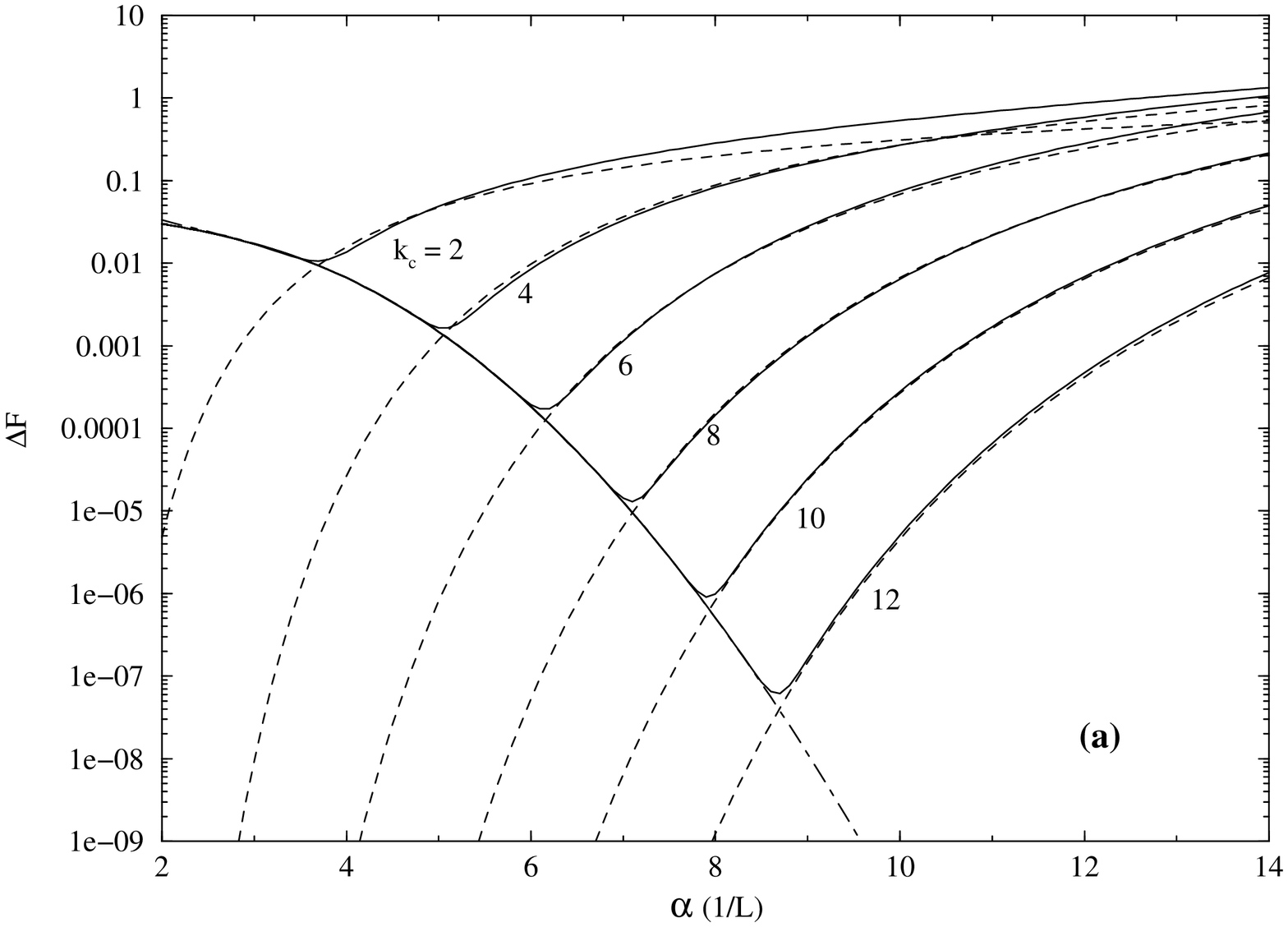}
  \includegraphics*[height=15.0 cm, angle=270]{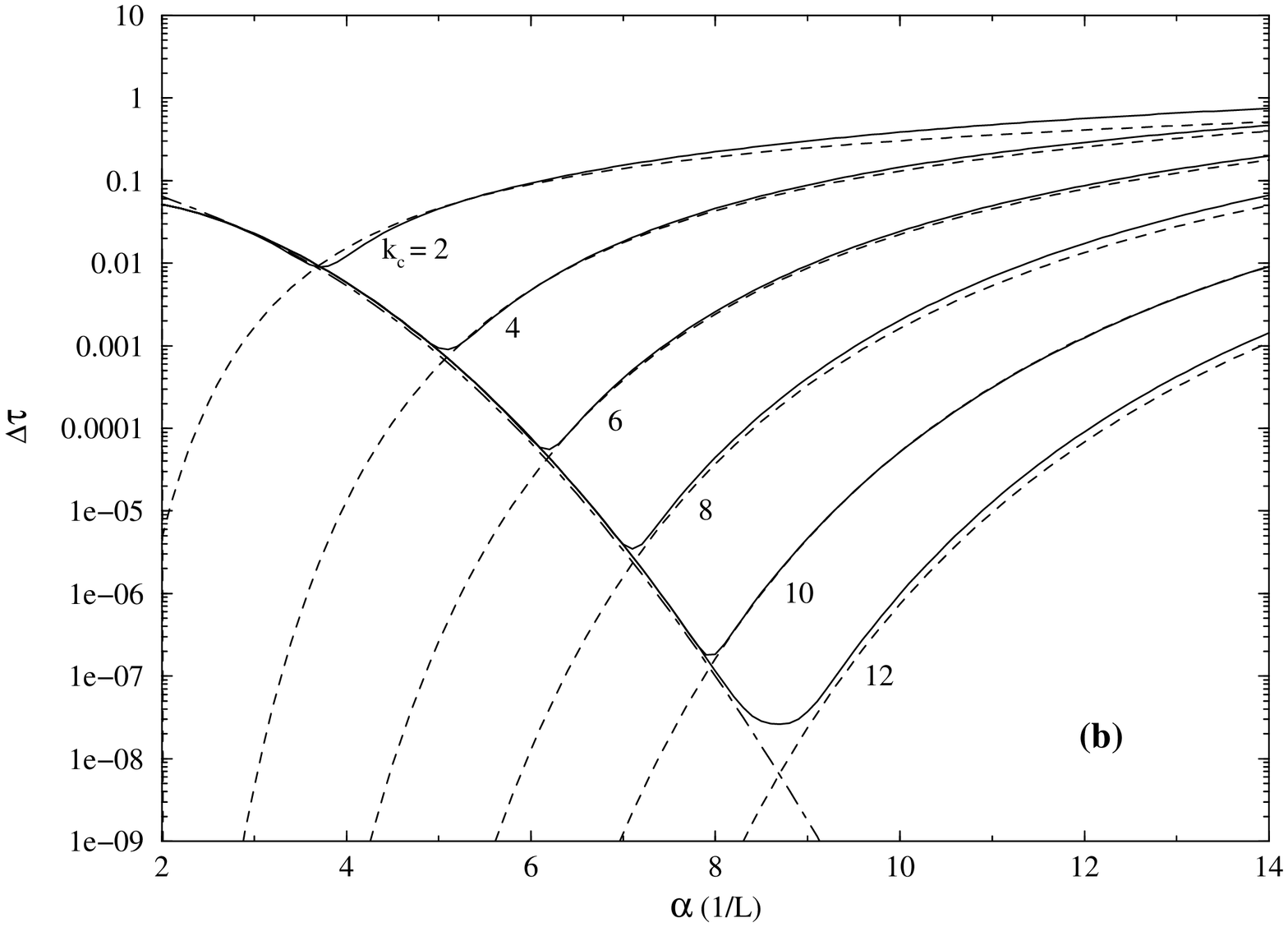}
  \end{center}
  \includegraphics*[height=15.0 cm, angle=270]{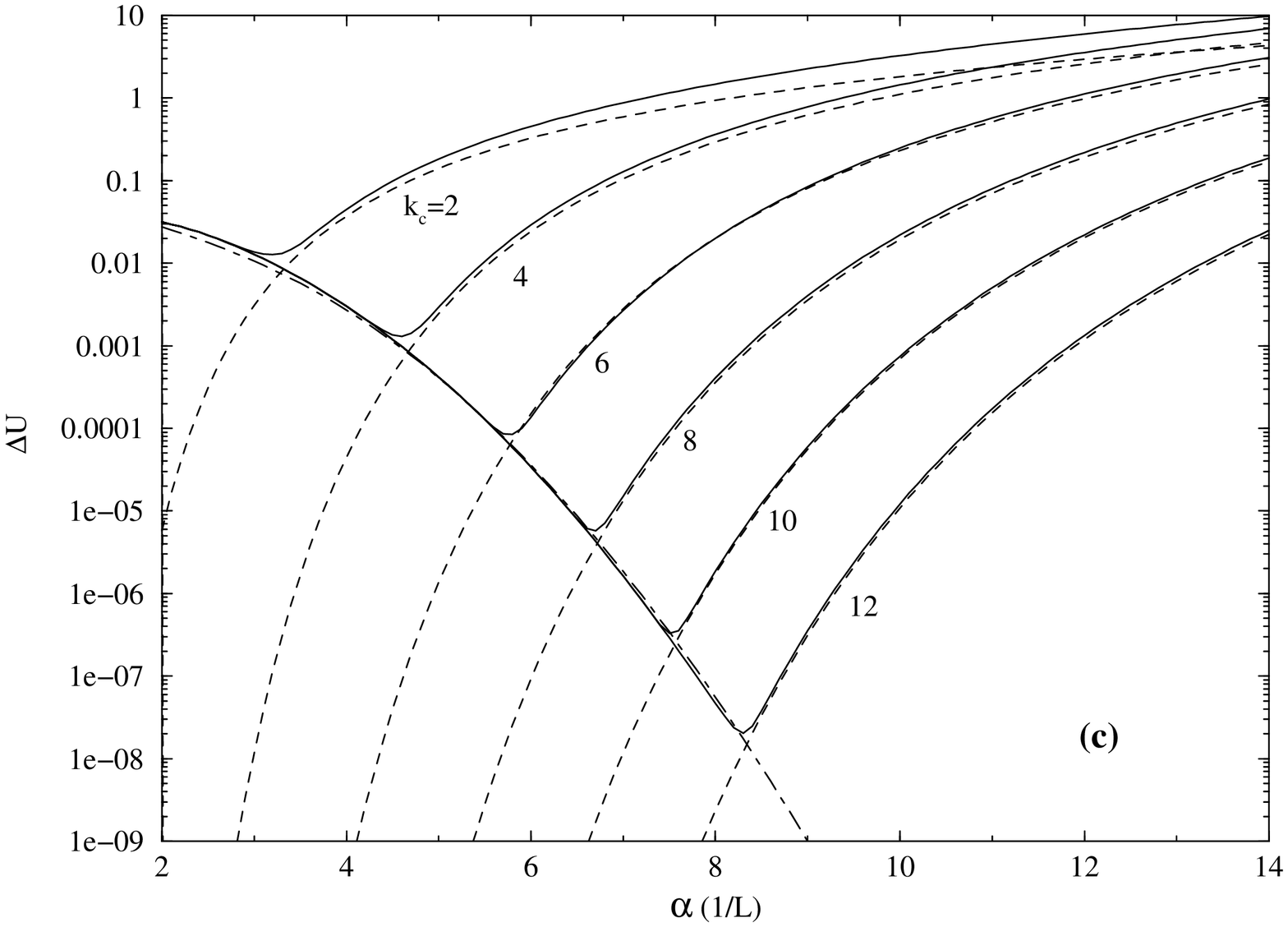}
  \caption{The rms errors (solid lines) on the forces $\Delta F$
 (a), torques $\Delta \tau$ (b) and total energy $\Delta U$
 (c) as a function of the splitting parameter $\alpha$. The results
 on $\Delta F$ and $\Delta \tau$ are obtained for the model
 system of 100 randomly distributed dipoles, while $\Delta U$ is 
 calculated from an average over 10 random configurations which
 are generated in the same way as the model system.
 The real-space cutoff is taken to be $r_c = 5.0$. The dot-dashed curves
 are the corresponding real-space error estimates achieved with 
 Eqn.(\ref{forcer}, \ref{torquer}) and (\ref{energyr}). The dashed
 curves are the k-space error estimates obtained with
 Eqn.(\ref{forcek}, \ref{torquek}) and (\ref{energyk}).
 The units of the energy, force and torque are 
 $\CALP^2/\CALL^3, \CALP^2/\CALL^4$ and $\CALP^2/\CALL^3$, respectively.} 
  \label{geometry} 
\end{figure}

\newpage

\begin{figure}[htb] 
\begin{center}
\includegraphics*[height=14.0 cm, angle=270]{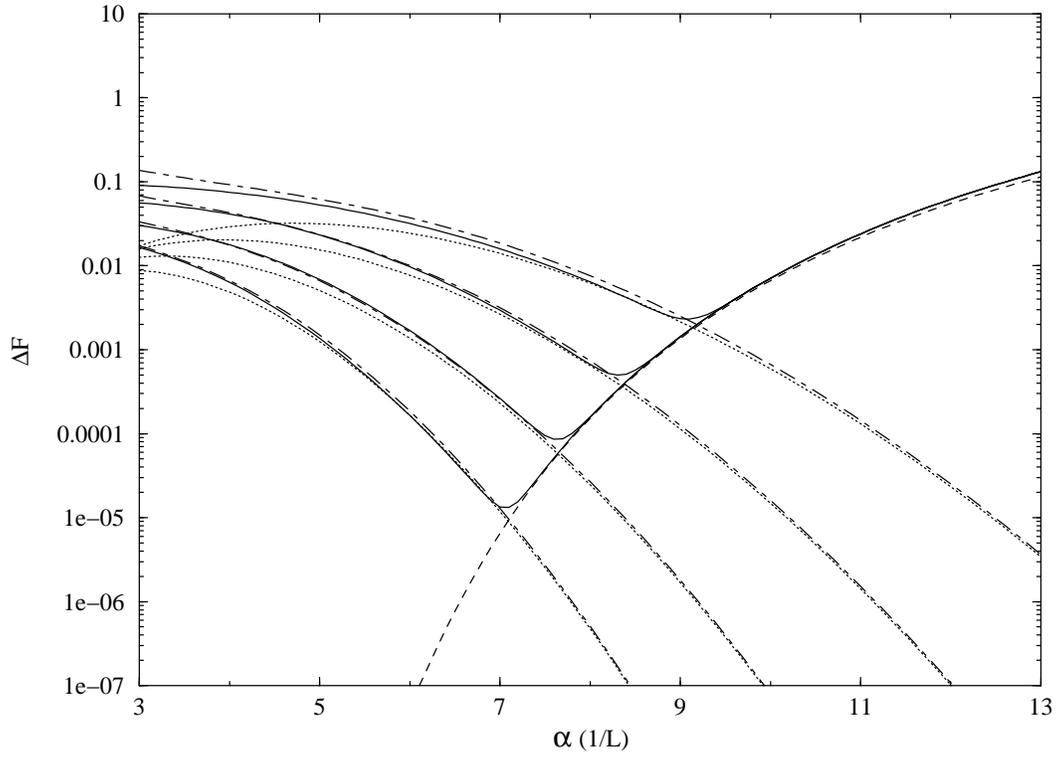}
  \end{center}
  \caption{The rms errors (solid lines) in the forces $\Delta F$
 for the model system. The k-space cutoff is taken to be $k_c = 8$.
 From top to bottom the real-space cutoff varies like $r_c = 3.0, 3.6,
 4.3, 5.0$. The dot-dashed curves are the corresponding real-space
 error estimates achieved with Eqn.(\ref{forcer}). The results
 obtained with the simplified formula Eqn.(\ref{forcers}) (the dotted curves)
 are also shown for comparison. The dashed curves are the k-space
 error estimates obtained with Eqn.(\ref{forcek}). The unit of the
 force is $\CALP^2/\CALL^4$.} 
  \label{geometry} 
\end{figure}

\begin{figure}[htb] 
\begin{center}
 \includegraphics*[height=14.0 cm, angle=270]{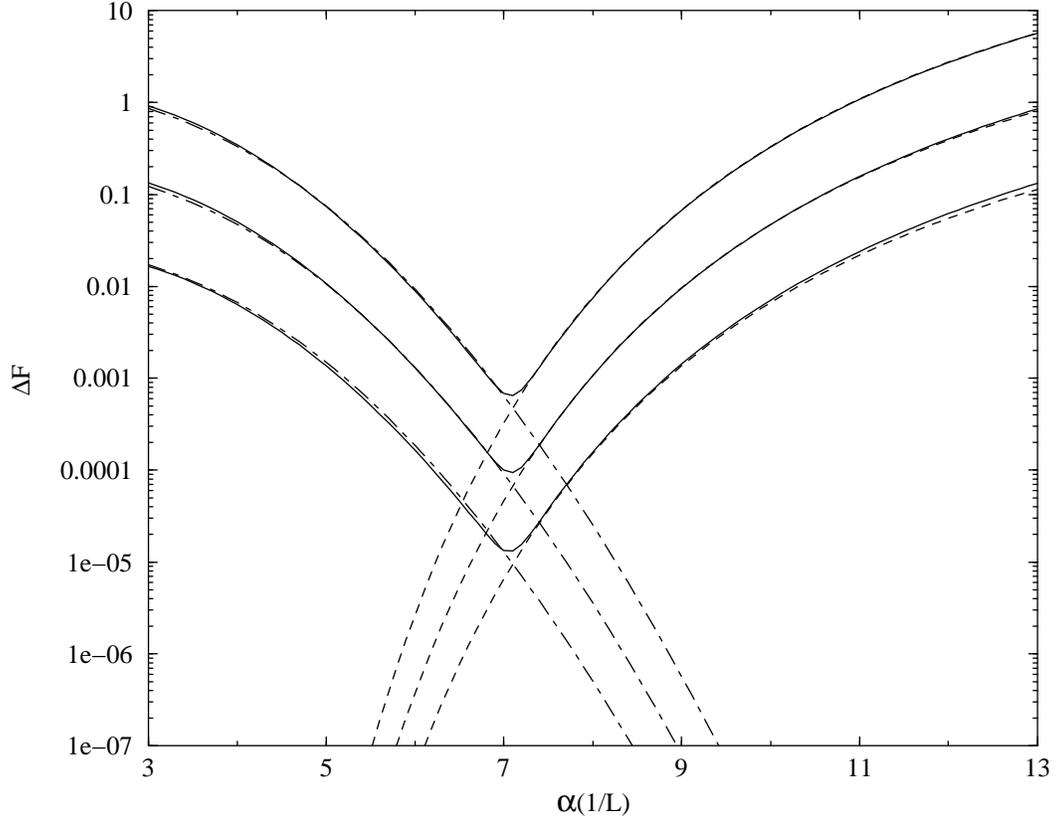}
\end{center}
\caption{Test of the $\CALM^2 N^{-1/2}$ scaling of $\Delta F$.
 The three solid lines show the rms error $\Delta F$ for systems
 which are different in their $(\CALM^2,N)$ values. From top
to bottom they are characterized by (10000, 400), (1000, 200) and
(100, 100) (the last one is the model system discussed in Fig.(1-2)).
The real and k-space cutoffs are $r_c = 5.0$ and $k_c = 8$, respectively.
The dot-dashed and dashed curves are the corresponding real-space and
k-space error estimates obtained with Eqn.(\ref{forcer}) and
(\ref{forcek}), respectively. The unit of the force is $\CALP^2/\CALL^4$.}
\label{geometry} 
\end{figure}

\begin{figure}[htb]
 \begin{center}
   \includegraphics*[height=8.0 cm, angle=270]{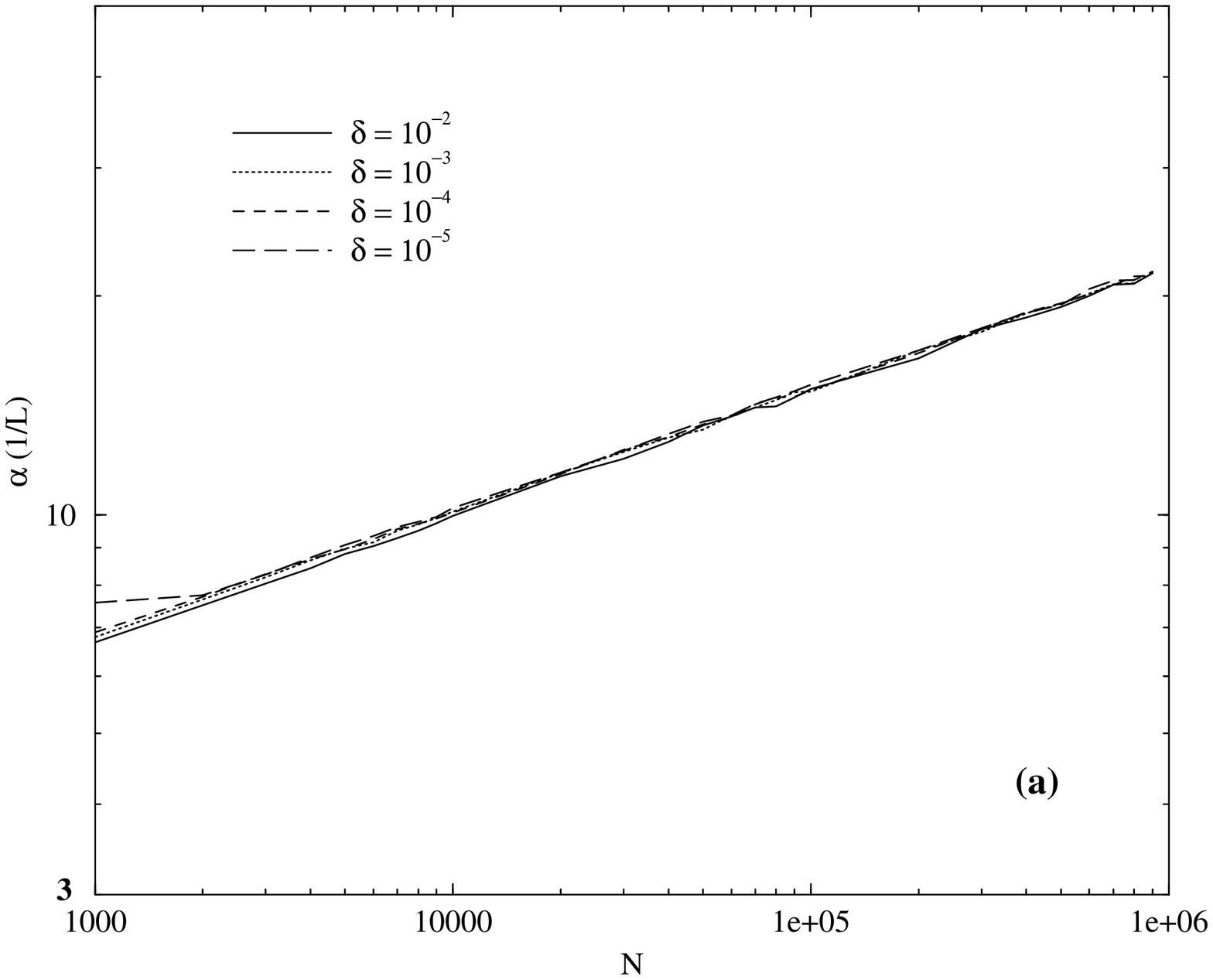}
  \includegraphics*[height=8.0 cm, angle=270]{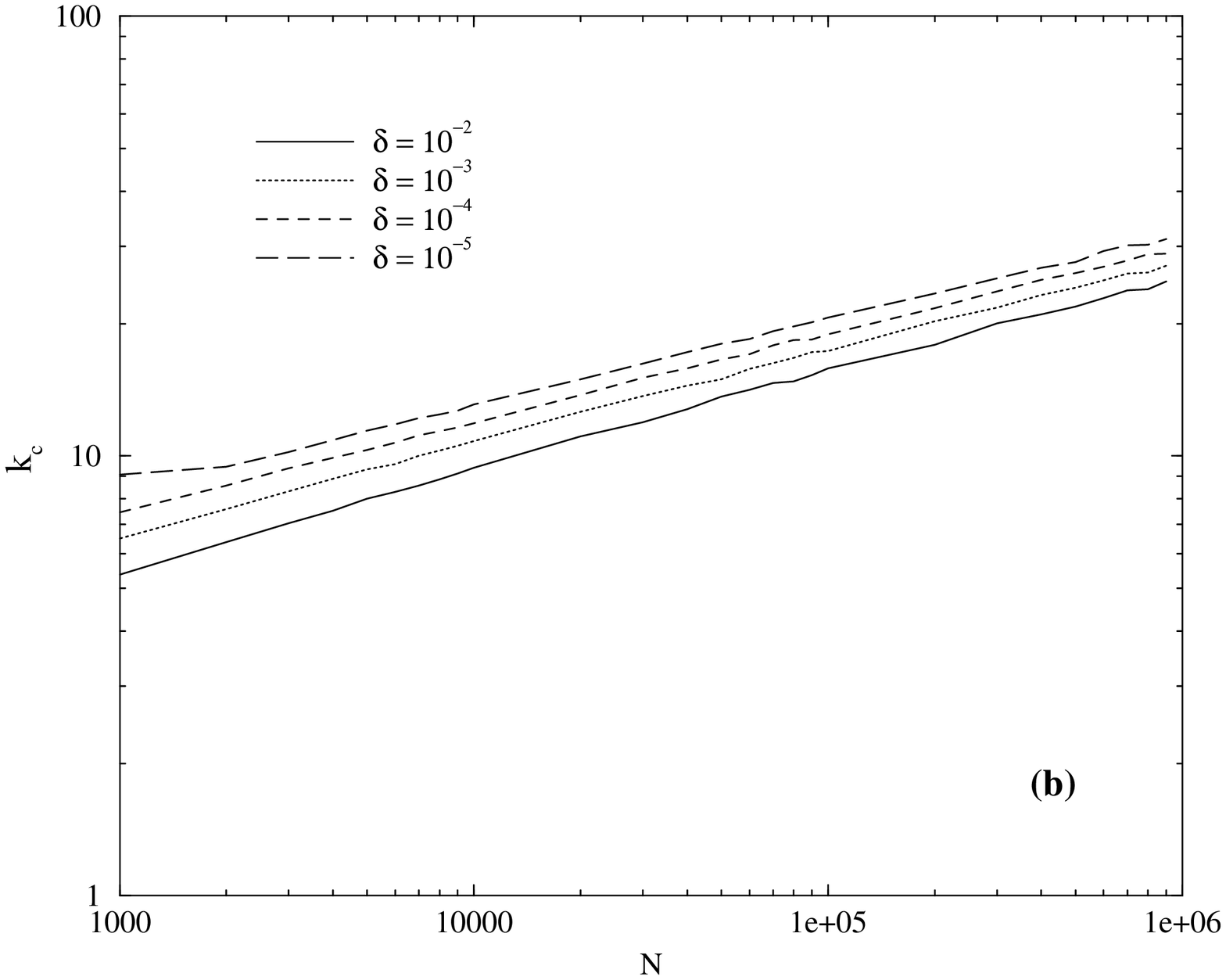} 
  \includegraphics*[height=8.0 cm, angle=270]{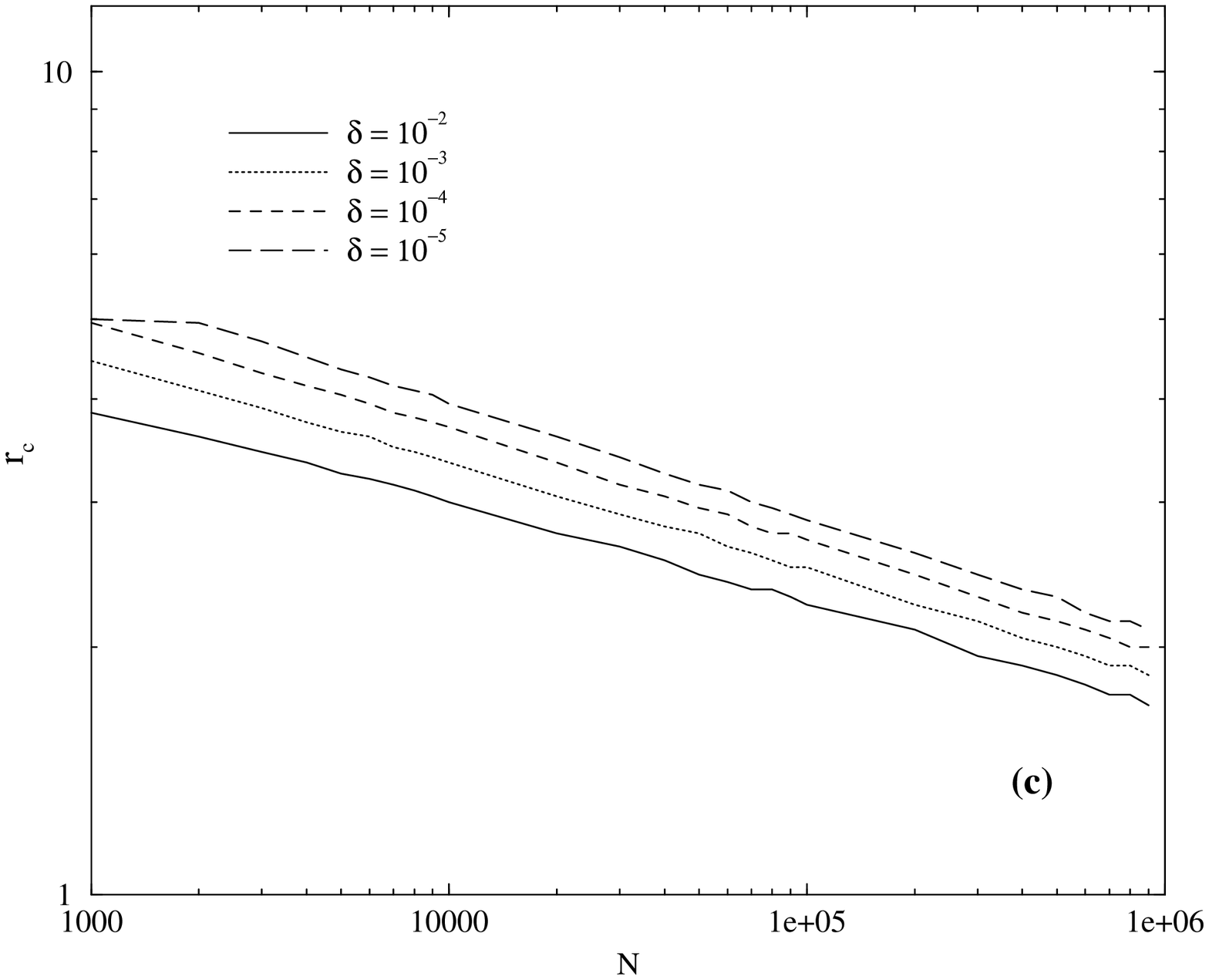}
  \includegraphics*[height=8.0 cm, angle=270]{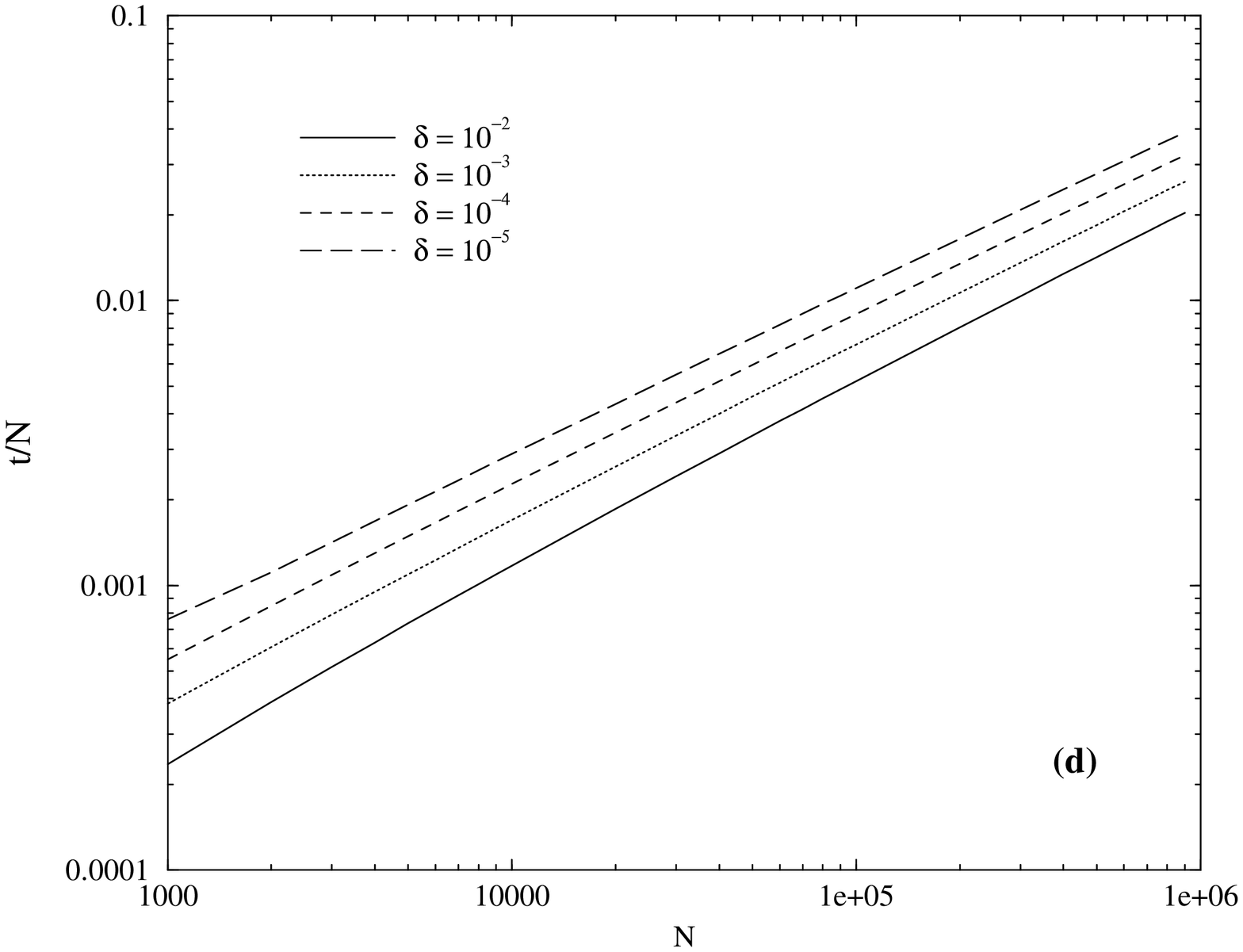}
 \end{center}
\caption{Optimal values of the parameters $\alpha$ (a), $k_c$ (b) and
  $r_c$ (c) as well as the corresponding minimized computation time
  $\CALT/N$ (d) as a function of the number of particles.}
\label{geometry} 
\end{figure}

\begin{figure}[htb] 
 \begin{center}
 \includegraphics*[height=14.0 cm, angle=270]{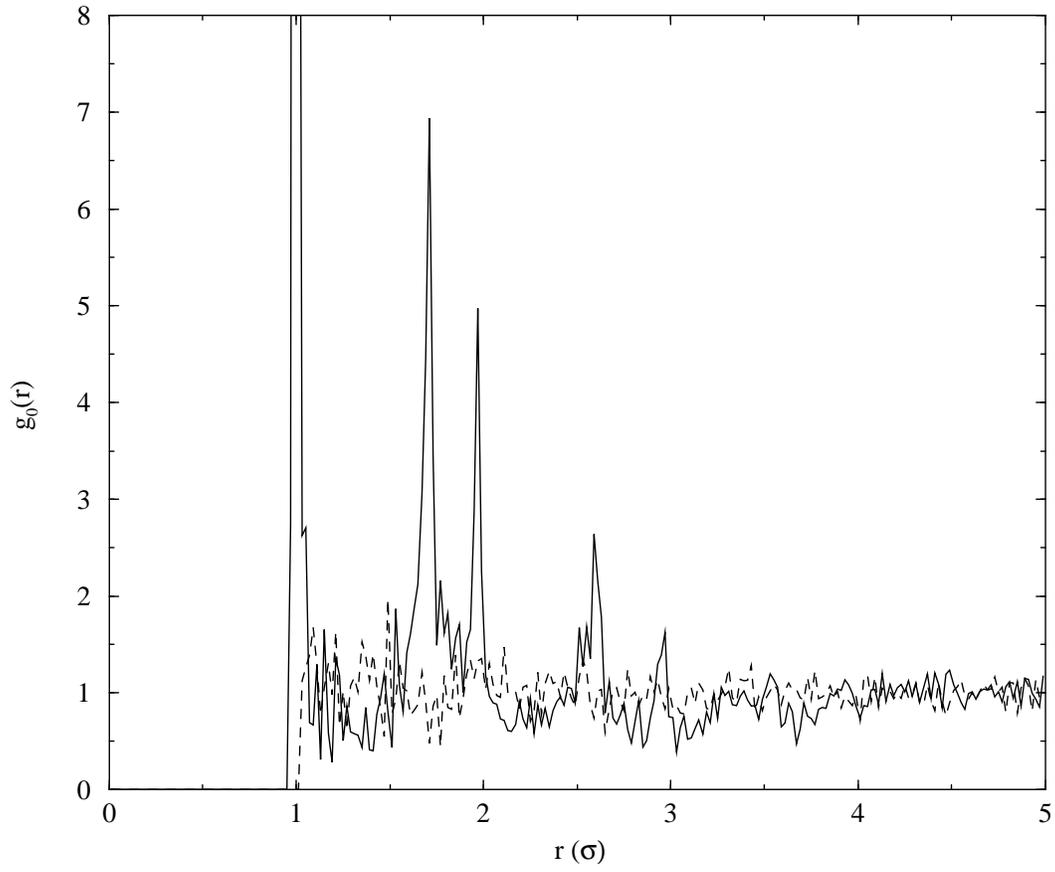}
 \end{center}
\caption{The radial distribution function (RDF) of the initial (dashed
line) and final (solid line) configurations of the electrorheological
system studied. The sharp peaks in the RDF of the final configuration
reflect the formation of chain-like structures along the field direction.}
\label{geometry} 
\end{figure}

\begin{figure}[htb]
 \begin{center}
\includegraphics*[height=14.0 cm, angle=270]{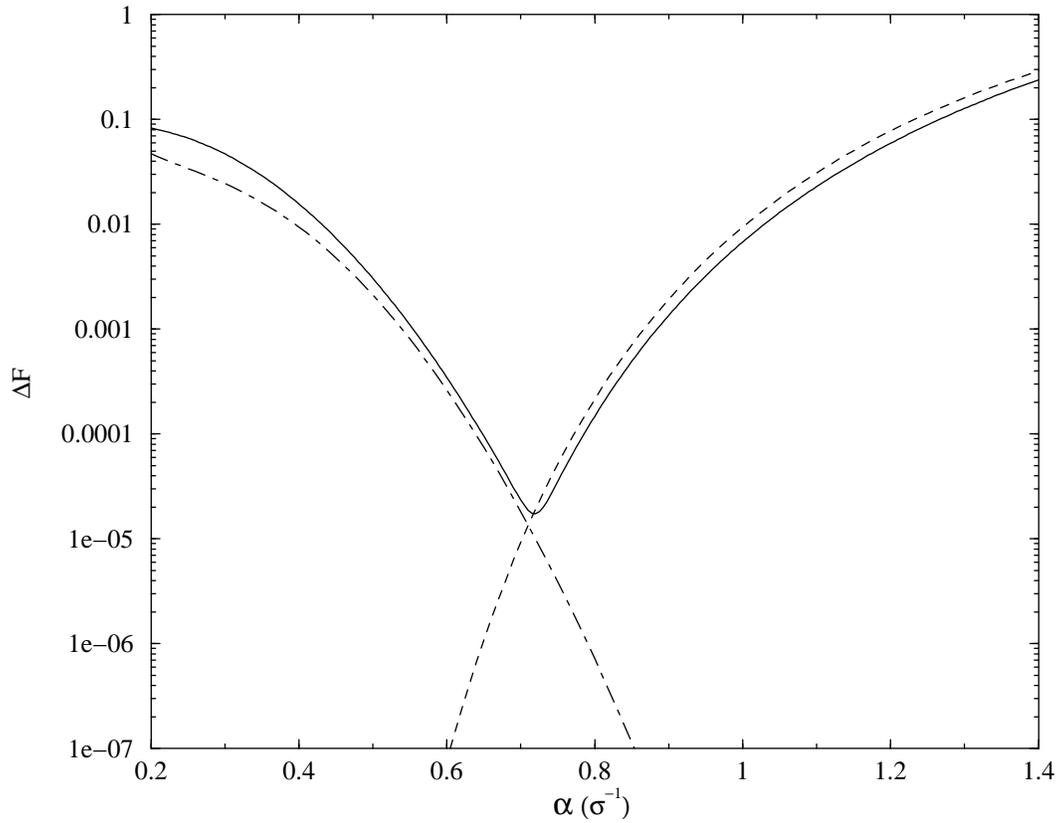}

 \end{center}
\caption{The rms force error $\Delta F$ for the final structure
of the electrorheological system studied (solid line). The dot-dashed
line is the error estimate for the real-space contribution
[Eqn.(\ref{forcer})] and the dashed line is that for the k-space
[Eqn.(\ref{forcek})]. The unit of the force is $\CALP^2/\sigma^4$.}
\label{geometry} 
\end{figure}
\end{document}